\documentclass[twocolumn,tighten]{aastex631}

\begin{document}

\shortauthors{Luhman}
\shorttitle{IMF of Stars and Brown Dwarfs in Upper Sco}

\title{The Initial Mass Function of Stars and Brown Dwarfs in the Upper Sco
Association\footnote{Based on observations made with the Gaia
mission, the Two Micron All Sky Survey, the Wide-field Infrared Survey Explorer,
the United Kingdom Infrared Telescope Infrared Deep Sky Survey, the Visible 
and Infrared Survey Telescope for Astronomy Hemisphere Survey, Pan-STARRS1,
the NASA Infrared Telescope Facility, Cerro Tololo Inter-American Observatory,
Gemini Observatory, and the European Southern Observatory.}}

\author{K. L. Luhman}
\affiliation{Department of Astronomy and Astrophysics,
The Pennsylvania State University, University Park, PA 16802, USA;
kll207@psu.edu}
\affiliation{Center for Exoplanets and Habitable Worlds, The
Pennsylvania State University, University Park, PA 16802, USA}

\begin{abstract}

I present infrared spectroscopy of 37 brown dwarf candidates in the Upper Sco
association, 35 of which are classified as young and cool, making them likely
members. This sample includes many of the faintest spectroscopically
confirmed members ($K=16$--17 mag), which should have masses down to 
$\sim0.007$-0.01 $M_\odot$ for the range of ages in Upper Sco (7--14 Myr).
Using my updated membership catalog for Upper Sco, I have estimated
the initial mass function (IMF) for a field in the center of 
the association that encompasses $\sim80$\% of the known members.
I have derived IMFs in the same manner for previous membership samples in
three other star-forming populations, consisting of IC 348, Taurus, and 
Chameleon I. When using logarithmic mass bins, the substellar
IMFs for Upper Sco and the other young regions are roughly flat down
to the completeness limits of $\sim$0.01~$M_\odot$.
These IMFs are broadly similar to mass functions recently measured
for the solar neighborhood.
Finally, I have used W1$-$W2 colors to search for excess emission
from circumstellar disks among the late-type objects in my new census of
Upper Sco. I measure an excess fraction of 52/200 for members 
with spectral types of M6.25--M9.5, which is similar to results 
from previous membership catalogs.
For the L-type members, it is difficult to detect the small W2 excess emission
produced by typical disks around brown dwarfs because of the large uncertainties
in spectral types, which preclude accurate estimates of the photospheric colors.
Thus, W2 photometry provides poor constraints on the presence of disks for
the L-type members of Upper Sco.

\end{abstract}

\section{Introduction}
\label{sec:intro}

The Upper Scorpius subgroup in the Scorpius-Centaurus (Sco-Cen) OB association 
has an average age of $\sim10$ Myr, a distance of $d\sim145$~pc, and $>$1000 
members, making it one of the richest nearby populations of newborn stars 
\citep{pm08}. As such, Upper Sco is an attractive site for measuring the 
initial mass function (IMF) down to low masses and with good statistical 
constraints. Because Upper Sco covers a large area of sky ($\sim100$ deg$^2$),
identifying its substellar population has been facilitated by wide-field 
surveys at optical and infrared (IR) wavelengths, including 
the Two Micron All Sky Survey \citep[2MASS,][]{skr06},
the Deep Near-Infrared Survey of the Southern Sky \citep[DENIS,][]{epc99},
the United Kingdom Infrared Telescope Infrared Deep Sky Survey
\citep[UKIDSS,][]{law07}, the Wide-field Infrared Survey Explorer
\citep[WISE,][]{wri10}, Pan-STARRS1 \citep[PS1,][]{kai02,kai10},
the Visible and Infrared Survey Telescope for Astronomy (VISTA)
Hemisphere Survey \citep[VHS,][]{mcm13}, and the Gaia mission 
\citep{per01,deb12,gaia16b}.
Numerous studies have used photometry and astrometry from those surveys
to identify brown dwarf candidates in Upper Sco
\citep{mar04,mar10,sle06,sle08,lod06,lod07,lod08,lod11a,lod11b,lod13,lod18,lod21,daw11,daw14,pen16,luh18,luh20,luh22sp}.
\citet{luh22sp} presented a compilation of $\sim$1700 sources
that have evidence of membership in Upper Sco from spectroscopy and astrometry,
which included 237 objects that have spectral types later than M6 and thus 
are likely to have substellar masses \citep{bar15}.

Based on their analysis of imaging data from several telescopes, \citet{mir22} 
(hereafter M22) reported the discovery of 
candidates for ``free-floating planets" in Upper Sco and the neighboring
Ophiuchus cloud, which had mass estimates below the deuterium burning 
limit (13 $M_{\rm Jup}$). The number of these objects ranged from 70 to 170
for ages of 3--10 Myr.
However, as discussed in this study, a majority 
of those sources had been identified as brown dwarf candidates in previous 
surveys, and half of them had been observed with spectroscopy in earlier work. 
In addition, I adopt definitions for brown dwarfs and planets that are 
based on their formation mechanisms such that substellar
objects forming in circumstellar disks are planets and those forming in 
a star-like manner are brown dwarfs \citep{cha07}, in which case none of 
the candidates qualify as free-floating planets since they lack evidence 
of formation within disks.

In the available imaging surveys of Upper Sco, there remain many viable
brown dwarf candidates that lack spectroscopy, which is needed for confirming
their youth and late spectral types. In this paper, I seek to improve
the completeness of the spectroscopic census of brown dwarfs in the central,
richest area of the association.  I then use the updated catalog of members
to estimate the mass function of stars and brown dwarfs down to masses of
$\sim0.01$ $M_\odot$ ($\sim10$~$M_{\rm Jup}$). 

\section{Selection of Brown Dwarf Candidates for Spectroscopy}
\label{sec:select}

\citet{luh18} identified candidates for stellar and substellar members of
Upper Sco using photometry and astrometry from the 2MASS Point Source Catalog 
(\dataset[doi:10.26131/IRSA2]{http://doi.org/doi:10.26131/IRSA2}), 
the third data release of DENIS, the science verification release
and data release 10 of UKIDSS, the AllWISE Source Catalog 
(\dataset[doi:10.26131/IRSA1]{https://doi.org/doi:10.26131/IRSA1}),
the sixth data release of VISTA VHS, the first data release of Gaia,
and the first data release of the 3~$\pi$ survey from PS1 \citep{cha16,fle16}.
\citet{luh20} and \citet[][hereafter L22]{luh22sc} updated the selection of 
candidates to
include the second and third data releases of Gaia (DR2 and DR3), respectively 
\citep{bro18,bro21,val23}. For this study, I have incorporated the CatWISE2020 
Catalog \citep[][]{eis20,mar21} 
(\dataset[doi:10.26131/IRSA551]{https://doi.org/doi:10.26131/IRSA551}) 
from the NEOWISE survey \citep{mai14},
which is based on the images from the WISE survey and several years of
additional observations in the W1 and W2 bands (3.6 and 4.5~\micron).
M22 made use of imaging from several telescopes in the
generation of a catalog for a large field surrounding Upper Sco and Ophiuchus,
so I have included it in my selection of brown dwarf candidates.

\begin{figure}
\epsscale{1.2}
\plotone{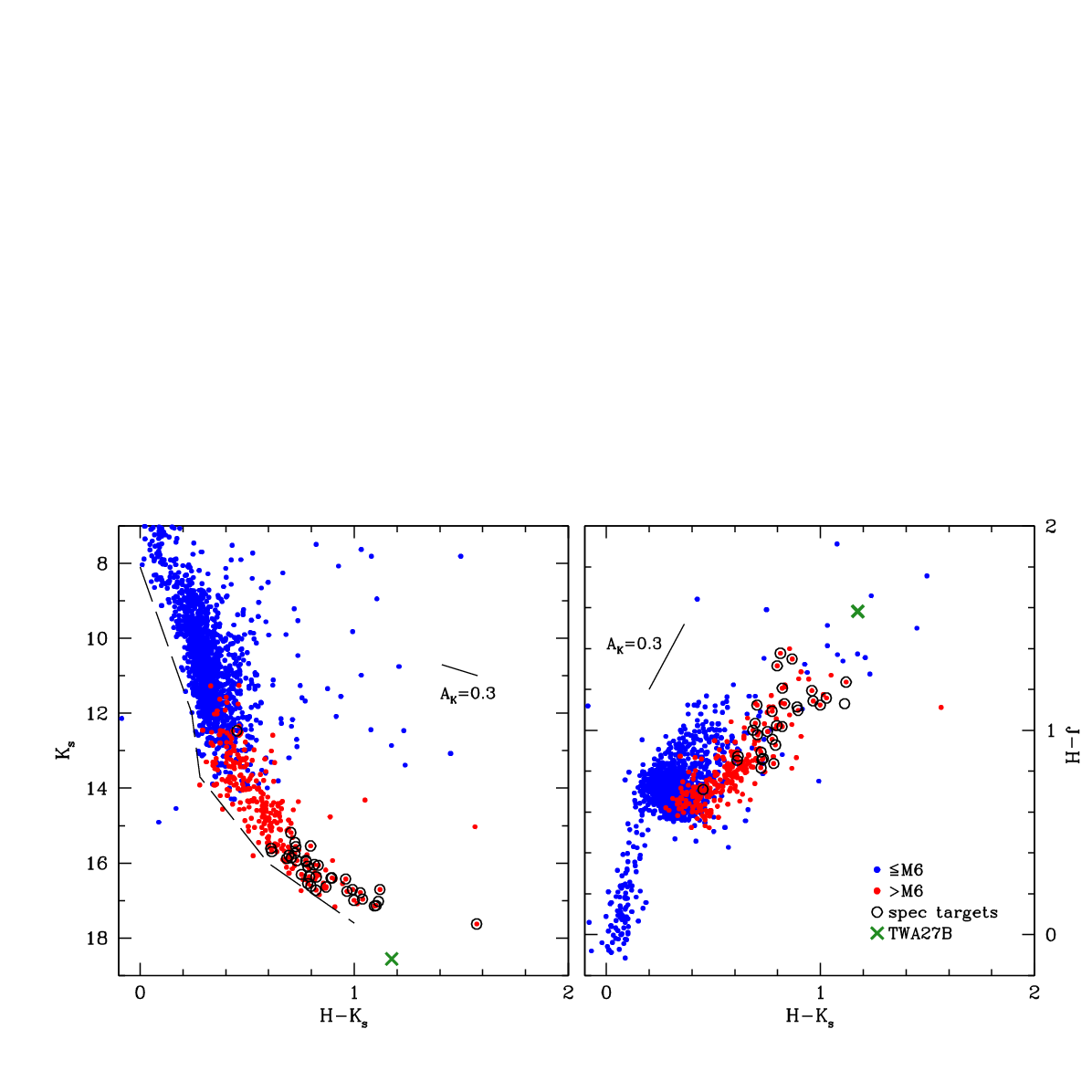}
\caption{Near-IR color-magnitude and color-color diagrams for 
adopted members of Upper Sco (Table~\ref{tab:usco}) (blue and red points)
and the brown dwarf candidates observed spectroscopically in this work
(circles).  Six candidates lack $J$ measurements, so they are absent from the 
diagram on the right. Two candidates were found to be nonmembers based on
the spectroscopy. The planetary mass companion TWA 27B is included for
a distance of 145 pc \citep[cross,][]{luh23}.}
\label{fig:cmd}
\end{figure}

I have identified brown dwarf candidates in Upper Sco using updated versions
of the proper motion and photometric diagrams presented in \citet{luh18}.
In Figure~\ref{fig:cmd}, I show two examples of these diagrams,
consisting of $K_s$ and $J-H$ versus $H-K_s$. The diagrams contain my
adopted members of Upper Sco (Section~\ref{sec:imf}) and 37 candidates
observed spectroscopically in this work, which consist of the most promising
candidates (i.e., satisfy the criteria from
several photometric diagrams) within a
triangular field defined by \citet{luh20} that encompasses the
central concentration of members. Six of the candidates lack $J$
measurements, so they are absent from the color-color diagram.
Two of the candidates are classified as nonmembers (a field dwarf and
a galaxy) based on spectroscopy (Section~\ref{sec:class}).

\section{Spectroscopic Observations}

\subsection{Brown Dwarf Candidates in Upper Sco}

I performed near-IR spectroscopy on 37 brown dwarf candidates in Upper Sco 
using SpeX \citep{ray03} at the NASA Infrared Telescope Facility (IRTF), 
the Gemini Near-Infrared Spectrograph \citep[GNIRS,][]{eli06} at the Gemini 
North telescope, and FLAMINGOS-2 \citep{eik04} at the Gemini South telescope. 

\subsection{Disk-bearing Star in Ophiuchus}

SpeX data were taken for ISO Oph 178 ([GY92] 371, Gaia DR3
6049152649744764160), which is a disk-bearing star in Ophiuchus \citep{bon01}
that lacks a previous spectral classification.

\subsection{Candidates for Low-mass Stars in Upper Sco from Gaia}

I have analyzed new and archival spectra of 37 candidates for low-mass stars in 
Upper Sco and other populations in Sco-Cen, which have been identified 
with astrometry and photometry from Gaia DR3 (L22).
I collected optical spectra of 13 of those 37 candidates with the Cerro 
Tololo Ohio State Multi-Object Spectrograph (COSMOS)\footnote{COSMOS 
is based on an instrument described by \citet{mar11}.} on the 4 m
Blanco telescope at Cerro Tololo Inter-American Observatory (CTIO).
Archival SpeX data are available for nine of the 37 candidates through
program 2017A107 (PI: Z. Zhang). I also have made use of archival optical 
spectra of 15 of the 37 candidates that were observed with the European Southern
Observatory (ESO) Faint Object Spectrograph and Camera \citep[EFOSC,][]{buz84} 
on the ESO New Technology Telescope (NTT) through program 0105.C-0283(A)
(PI: H. Bouy).

The instruments and observing modes for the new (51)
and archival (24) spectra
are summarized in Table~\ref{tab:log}. The instruments and observing 
dates for individual targets are provided in Table~\ref{tab:spec}.

\subsection{Data Reduction}

\begin{figure}
\epsscale{1.3}
\plotone{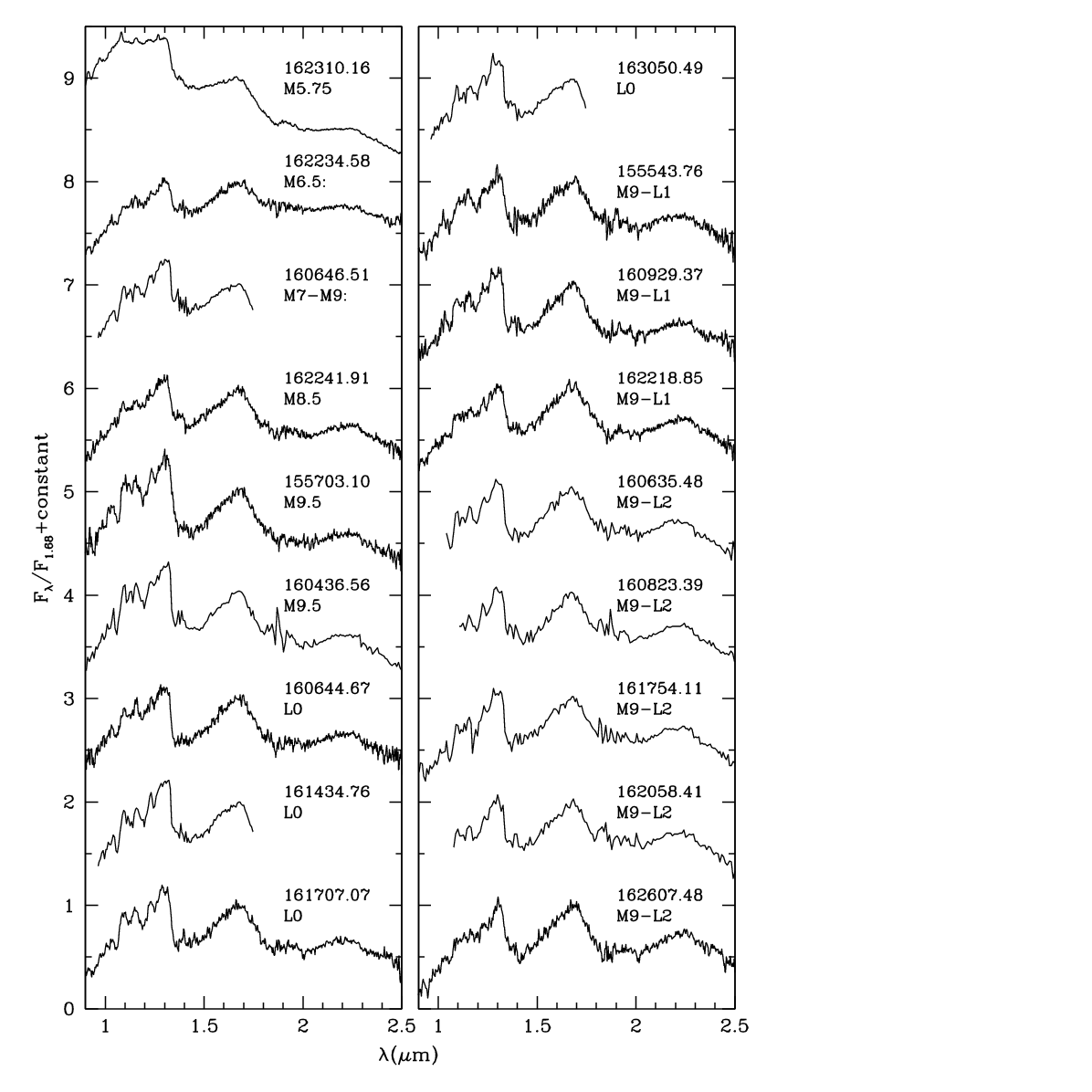}
\caption{Near-IR spectra of brown dwarf candidates in Upper Sco. 
The Gemini data have been binned to the resolution of the SpeX data.}
\label{fig:spec1}
\end{figure}

\begin{figure}
\epsscale{1.3}
\plotone{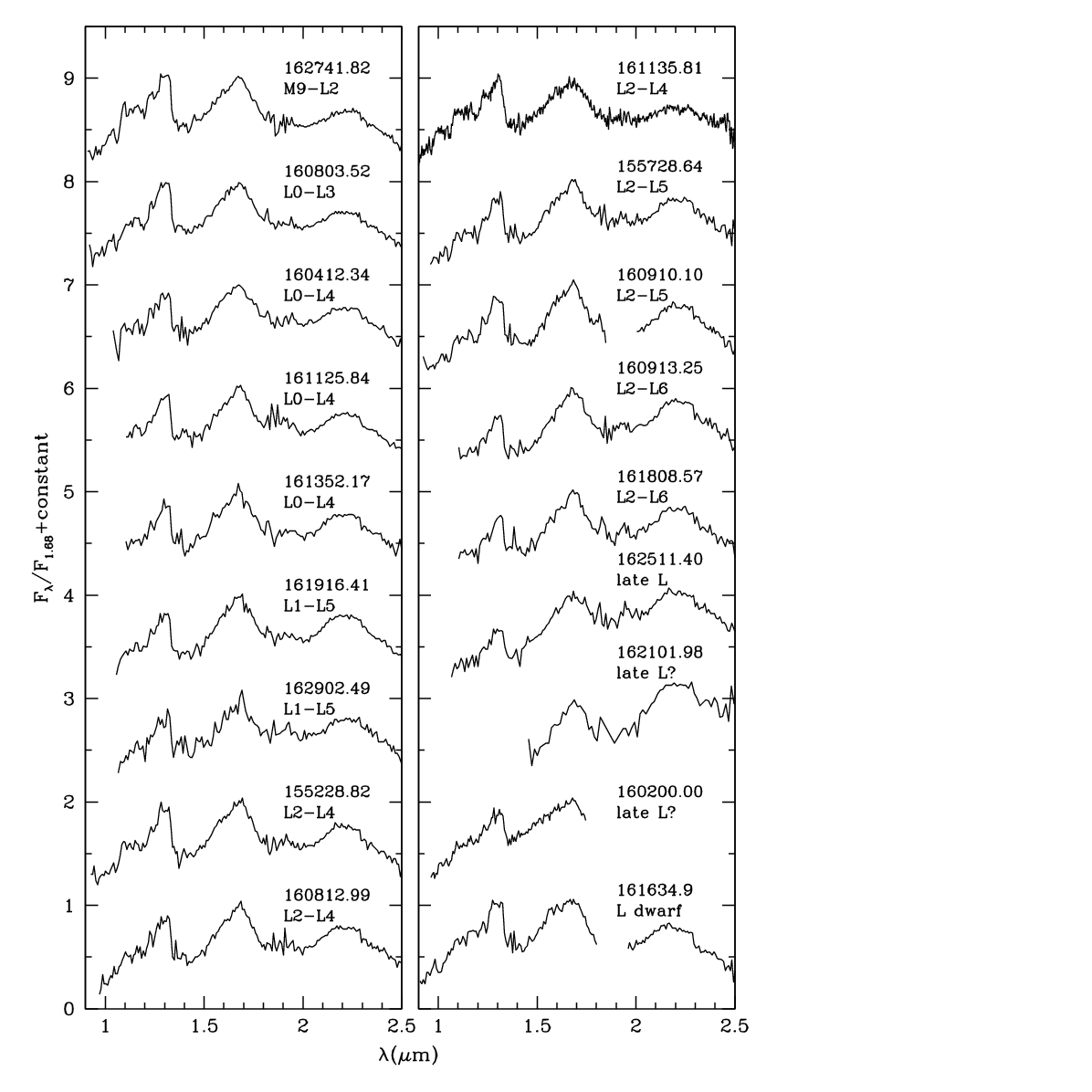}
\caption{Near-IR spectra of brown dwarf candidates in Upper Sco.
The Gemini data have been binned to the resolution of the SpeX data.}
\label{fig:spec2}
\end{figure}

The SpeX data were reduced with the Spextool package \citep{cus04},
which included a correction for telluric absorption \citep{vac03}.
The data from the remaining instruments were reduced with routines
within IRAF. The reduced spectra for the brown dwarf candidates are
presented in Figures~\ref{fig:spec1} and \ref{fig:spec2} with the exception
of the candidate classified as a galaxy. The spectra from COSMOS, SpeX, 
GNIRS, and FLAMINGOS-2 are available in an electronic file associated
with Figure~\ref{fig:spec1}.

\section{Spectral Classifications}
\label{sec:class}

I have assessed whether each of the spectroscopic targets has a young
age that is consistent with membership in Sco-Cen using Li and 
gravity-sensitive features like Na, FeH, and the $H$-band continuum
\citep[][L22]{luc01}. Most of the objects (68/75)
have evidence of youth in their spectra. The optical spectra for the young
sources have been classified through visual comparison
of molecular absorption bands (e.g., TiO, VO, H$_2$O) to
those of field dwarf standards for $<$M5 \citep{hen94,kir91,kir97} and
averages of dwarf and giant standards for $\geq$M5 \citep{luh97,luh99}.
I have classified the IR spectra of young sources through comparison to
standard spectra for young stars and brown dwarfs \citep{luh17}.
The measured spectral types and a flag for youth are included in 
Table~\ref{tab:spec}.

Among the 37 brown dwarf candidates observed spectroscopically, one is
a galaxy based on redshifted emission lines, 
one is a field L dwarf, and the remaining candidates have
types of late M or L and have evidence of youth, although a few of the
latter exhibit peculiarities in their spectra.
\citet{lod18} and \citet{bou22} obtained spectra of UGCS J160200.00$-$205734.0,
but the classifications were uncertain because of low signal-to-noise ratios
(S/Ns). \citet{lod18} detected strong H$\alpha$ emission that
suggested the presence of an accretion disk, but it lacks the mid-IR photometry 
($>$5~\micron) needed for confirmation of a disk.
My new spectroscopy has higher S/Ns than the previous data and shows
the triangular $H$-band continuum expected for young late-type objects.
It resembles young late L dwarfs in the strength of its steam absorption,
but it has a bluer slope (Figure~\ref{fig:spec2}).
Two objects, UGCS J162234.58$-$213905.0 and UGCS J160646.51$-$191703.3,
are unusually faint for members of Upper Sco with their spectral types.
Their spectral slopes also imply higher extinctions ($A_K\sim0.5$ and 0.25) 
than exhibited by most members of Upper Sco ($A_K<0.1$).
One possible explanation for these characteristics is that 
the two sources have
edge-on disks and are observed primarily in scattered light. Both of them
lack photometry at $>$5~\micron.

\citet{bou22} presented spectra of 18 brown dwarf candidates from M22,
four of which had been observed spectroscopically in
previous works \citep{lod18,luh20}.
Six of the targets from \citet{bou22} are also in my spectroscopic sample,
one of which was mentioned already (UGCS J160200.00$-$205734.0).
My spectra have higher S/Ns and broader wavelength coverage than those earlier
data.
For the nine candidates from \citet{bou22} that I have not classified in this
work or in previous studies, I have measured spectral types using the spectra
from \citet{bou22}. For two of those objects, DANCe J16255679$-$2113354 and 
DANCe J15582895$-$2530319, I find that the S/Ns are too low to determine
if they are young. My classifications for the remaining eight targets from
\citet{bou22} are included in the compilation of adopted members of Upper Sco 
in Section~\ref{sec:imf}.

\section{Initial Mass Function}
\label{sec:imf}

\subsection{IMF Sample}

Some recent studies of the membership of Upper Sco have been based
primarily on data from Gaia and have not considered the available
spectroscopic data \citep[e.g.,][]{mir22b}. 
However, spectroscopy is important for measurements of
spectral types (which are relevant to mass estimates) and confirmation of 
youth for candidate members, particularly for faint, low-mass objects 
that lack high precision astrometry from Gaia. Fortunately, extensive
spectroscopy has been performed on candidate members of Upper Sco,
as mentioned in Section~\ref{sec:intro}.

\citet{luh22sp} compiled stars in Upper Sco and Ophiuchus
that have spectral classifications, evidence of youth, and astrometry that
is consistent with membership. I have generated an updated compilation for
Upper Sco, which consists of stars that have (1) spectral classifications and 
evidence of youth (see Section~\ref{sec:class}), 
(2) locations within the boundary of Upper Sco from 
\citet{dez99} ($l=343$--$360\arcdeg$, $b=10$--$30\arcdeg$) and outside of 
the boundary for the Ophiuchus complex from \citet{esp18}, 
(3) Gaia DR3 parallaxes and proper motions that satisfy the membership
criteria from L22 if those measurements are available, and 
(4) non-Gaia proper motions that are consistent with membership
if Gaia measurements are not available. These criteria produce
a catalog that contains
1753 sources, 274 of which are later than M6 and thus are likely to be
substellar. This is the largest sample of probable brown dwarfs with spectral 
classifications in any young stellar population. The catalog is presented
in Table~\ref{tab:usco}, which includes all available spectral classifications.

\begin{figure}
\epsscale{1.2}
\plotone{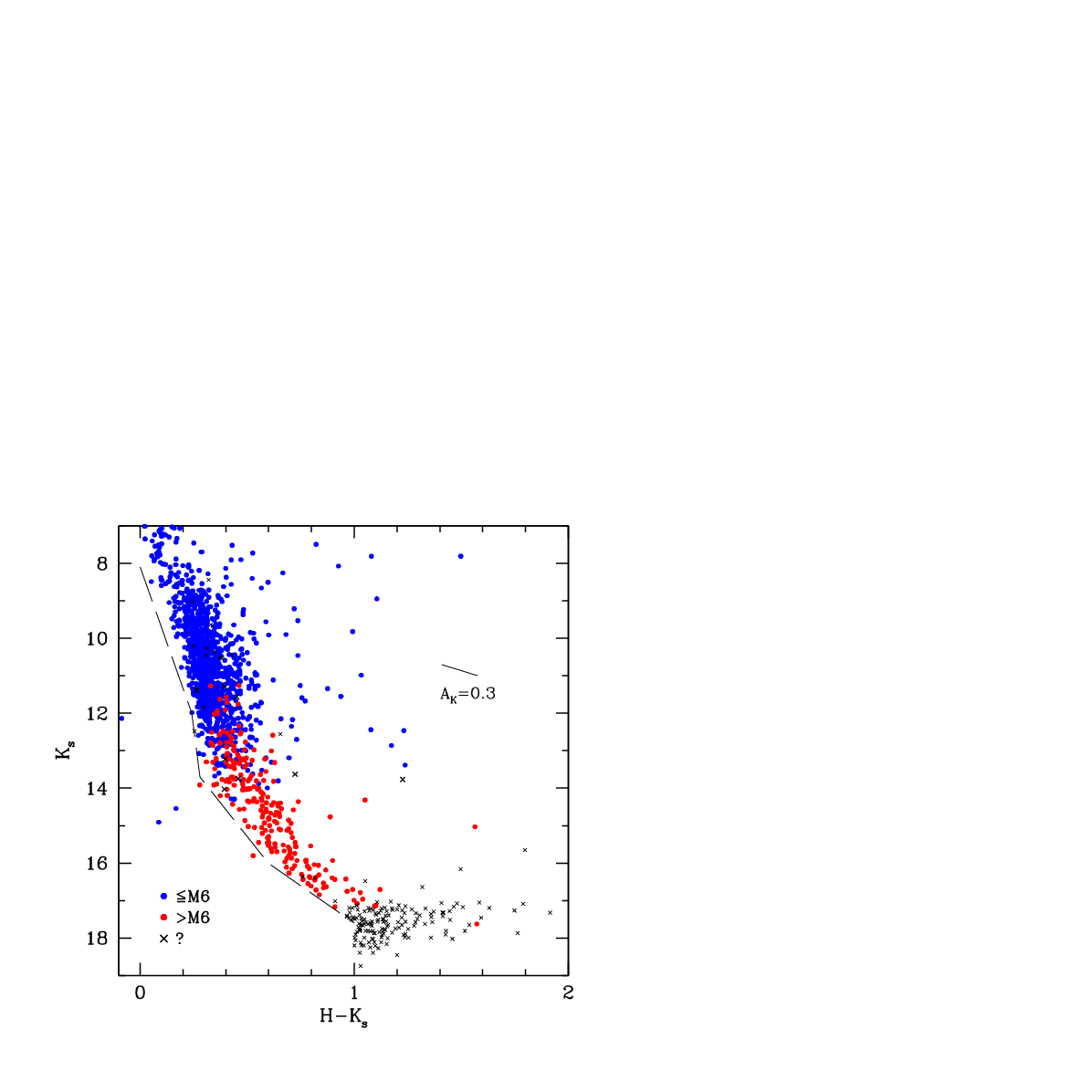}
\caption{
Near-IR color-magnitude diagram for members of Upper Sco that have spectral 
classifications and that are located within the 
triangular field from \citet{luh20}, which encompasses the central 
concentration in the association (red and blue points, Table~\ref{tab:usco}).
I also include the sources in that field that are not rejected by
available membership constraints (crosses). Most of the
latter sources have few bands of photometry available and are likely
to be background stars and galaxies.}
\label{fig:cmd2}
\end{figure}

As discussed in \citet{luh20} and \citet{luh22sc}, the Upper 
Centaurus-Lupus/Lower Centaurus-Crux (UCL/LCC) subgroup in Sco-Cen
overlaps with Upper Sco on the sky. Members of Upper Sco and UCL/LCC can be
readily distinguished from each other with the high precision astrometry 
from Gaia, but those data are not available for members later than $\sim$M7.
As a result, it is 
likely that my membership catalog for Upper Sco contains contamination from 
UCL/LCC at the latest spectral types. To minimize that contamination,
I estimate the IMF using adopted
members from Table~\ref{tab:usco} that are within the triangular field defined 
by \citet{luh20}, which encompasses the central concentration of stars
in Upper Sco, and outside the boundary of Ophiuchus from \citet{esp18}.
In Figure~\ref{fig:cmd2}, I have plotted a diagram of $K_s$ versus $H-K_s$ 
that contains the known members from Table~\ref{tab:usco} within that field
and all other IR sources that are not rejected by available
membership constraints (i.e., color-magnitude diagrams, proper motions, 
spectroscopy). Figure~\ref{fig:cmd2} indicates that my membership catalog for 
the central field has a high level of completeness ($>$90\%) down to $K_s=17$,
which corresponds to masses of $\sim0.007$--0.01~$M_\odot$ for a distance 
of 145 pc and ages of 7--14 Myr (Section~\ref{sec:age}) according to
evolutionary models \citep{bar15,cha23}. Thus, the lowest mass at which
members at all ages are detected (i.e., the mass completeness limit)
should be $\sim$0.01~$M_\odot$.

\subsection{Comparison to the Membership Catalog from M22}

M22 identified a sample of 3455 candidate members of Upper Sco and Ophiuchus
and estimated an IMF from that sample. They reported that $\sim$800 candidates
were newly identified, including those with mass estimates
below 13~$M_{\rm Jup}$, the number of which ranged between 70 and
170 for assumed ages of 10 and 3 Myr, respectively.
However, among those 170 objects below 13~$M_{\rm Jup}$ for 3 Myr, 
114 had been identified as brown dwarf candidates in previous studies,
84 of which had spectral classifications in earlier work 
\citep{bej08,lod08,lod13,daw14,chi20,luh18,luh20,esp20,luh22sp}.
My new catalog of adopted members with spectral classifications now contains
103 of those 170 candidates from M22 (93 of the 114 aforementioned
candidates from previous studies). Among the remaining 67 candidates,
$\sim$20 are viable candidates based on my analysis of their photometry and 
astrometry. They have $K_s\sim14.5$--16.5 mag and are located primarily in the
outskirts of Upper Sco, which is a reflection of the fact that my spectroscopy
has focused on candidates in the center of the association.
Meanwhile, among my 1753 adopted members, 128 are absent from the catalog 
of candidates in M22, 49 of which are later than M6.

\begin{figure}
\epsscale{1.2}
\plotone{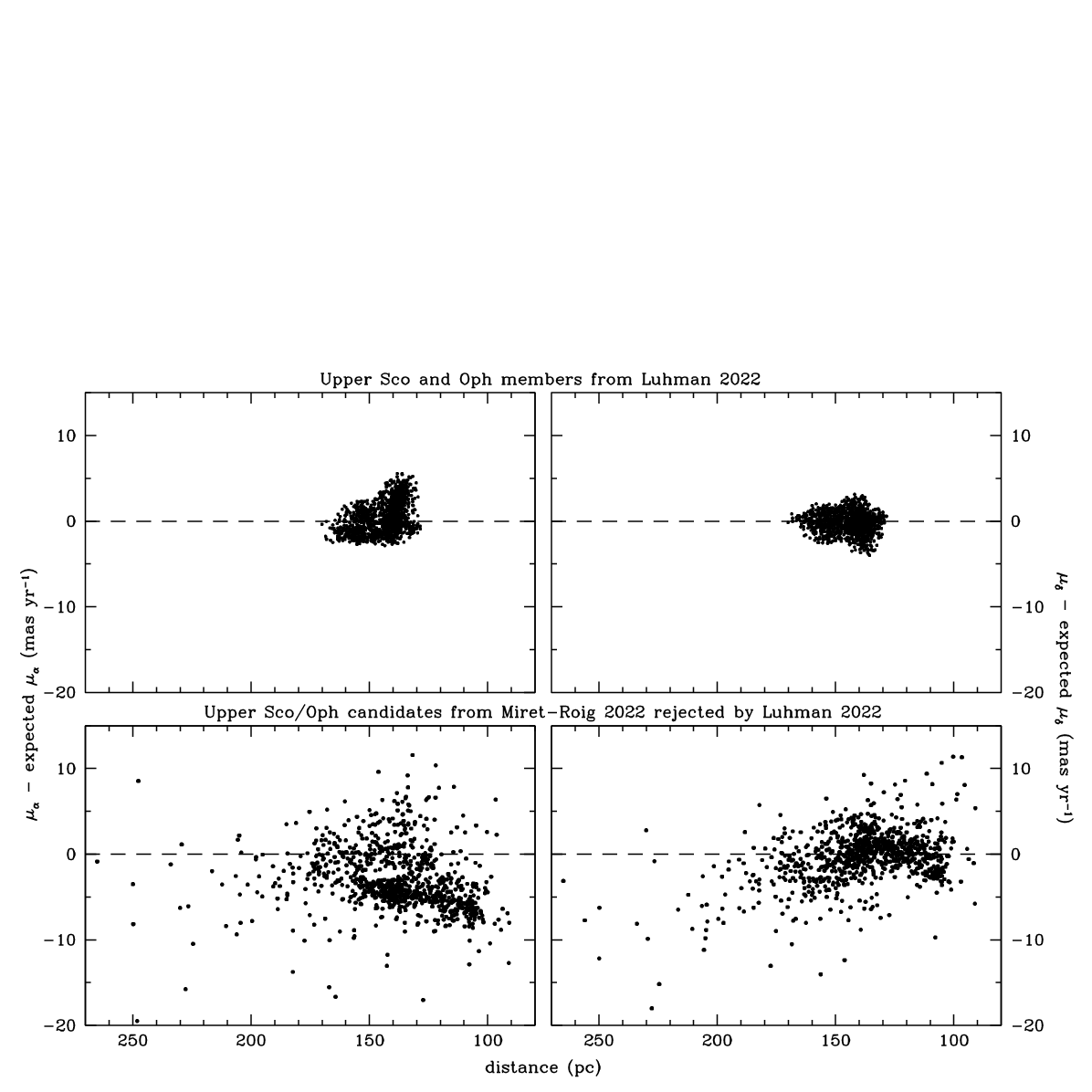}
\caption{Proper motion offsets versus parallactic distance based on Gaia DR3
for members of Upper Sco and Ophiuchus from \citet{luh22sc} (top) and candidate
members from \citet{mir22} that are rejected by the criteria in
\citet{luh22sc} (bottom).}
\label{fig:pp}
\end{figure}

More than 900 of the Gaia-detected candidates from M22 do not
satisfy the kinematic and photometric criteria for membership in
Upper Sco or Ophiuchus from L22. To illustrate that point,
I have plotted proper motion offsets\footnote{Proper motion offset
is defined as the difference between the observed proper motion of a
star and the motion expected at the celestial coordinates and parallactic
distance of the star for the median space velocity of Upper Sco.}
and parallactic distances in Figure~\ref{fig:pp} for members of Upper
Sco and Ophiuchus from L22 (top panel) and the M22 candidates that are
rejected by the criteria in L22 (bottom panel). 
The latter study demonstrated that members of Upper Sco
exhibit a tight clustering in their kinematics, which is illustrated in
the top panel of Figure~\ref{fig:pp}. In contrast, the rejected M22 candidates 
have a wide range of kinematics, which indicate that they are a mixture of
field stars and members of other populations in Sco-Cen, primarily UCL/LCC.

\subsection{Adopted Age}
\label{sec:age}

The age of Upper Sco is relevant to the estimation of masses for its
members and the construction of an IMF. Based on membership samples that were
available prior to Gaia, the main-sequence turnoff and the luminosities of
G/F stars in Upper Sco suggested an age of $\sim$10 Myr \citep{pec12,pec16}
while the positions of the low-mass stars in the Hertzsprung-Russell (H-R) 
diagram indicated ages of $\sim5$ Myr according to standard isochrones 
\citep{pre02,her15,pec16}, although models that included the inhibition of 
convection by magnetic fields could produce older ages \citep{fei16,mac17}.
Since those studies, the high precision astrometry and photometry from Gaia 
has improved the completeness and reliability of membership samples in Upper 
Sco and has provided more accurate measurements of the sequence of members in
the H-R diagram. For instance, \citet{luh20} estimated an age of $\sim$11 Myr
by combining the offset between Upper Sco and the $\beta$ Pic Moving Group 
(BPMG) in the H-R diagram with a lithium depletion age for BPMG and the 
change in luminosity with age predicted by evolutionary models. 

\citet{mir22b} divided Gaia-selected members of Upper Sco and 
Ophiuchus into seven populations based on their clustering in kinematics and 
spatial positions. These populations exhibited a range of ages in Gaia 
color-magnitude diagrams (CMDs).
I have compared the CMDs for those groups to the CMD for the TW Hya 
association (TWA), which has an expansion age of $\sim$10 Myr \citep{luh23twa}.
Among low-mass stars ($G_{\rm BP}-G_{\rm RP}=1.4$--2.8), I estimate the 
following offsets in $M_{G_{\rm RP}}$ relative to TWA for five of the groups, 
where a positive value corresponds to an older age for the latter: 
$\sim$0.2 for $\alpha$ Sco, no offsets for $\beta$ Sco and $\sigma$ Sco, 
$\sim-0.05$ for $\delta$ Sco, and $\sim-0.2$ for $\nu$ Sco.
The range of offsets corresponds to 7--13 Myr for the luminosity evolution
predicted by theoretical models \citep{bar15,cho16,dot16,fei16}
and 7--14 Myr for the evolution observed between TWA and the 32 Ori 
association \citep{luh25}.
These results are consistent with the average age of $\sim$10 Myr that has
been estimated for Upper Sco in some of the studies mentioned earlier.
Since most substellar members of Upper Sco lack the Gaia astrometry
needed for assigning them to individual groups within Upper Sco, I adopt
an age range of 7--14 Myr when estimating their masses for the IMF.

One of the two remaining groups from \citet{mir22b}, $\pi$ Sco, has distinct 
kinematics from the other Upper Sco groups and has been classified as part of 
UCL/LCC (L22). The final group has the youngest age based on its CMD, exhibits 
the largest reddenings, and is concentrated near the Ophiuchus clouds. 
A majority of those stars were classified as members of the Ophiuchus complex
in L22.  \citet{mir22b} named that group after $\rho$ Oph. However, that star
is a member of a small, compact group that is older than the stars associated
with the Ophiuchus clouds \citep{pil16,esp20}, so $\rho$ Oph is not an 
appropriate name for the stars in those clouds.

\subsection{Construction of IMF}
\label{sec:mass}

In previous studies of star-forming regions and young associations, I have 
characterized their IMFs in terms of the distributions of observational 
parameters that are related to stellar mass (spectral type and IR magnitudes)
to avoid the uncertainties in deriving masses of young stars with evolutionary 
models. However, I wish to compare the IMF for Upper Sco to the mass function 
in the solar neighborhood to check if they agree
\citep{bes24,kir24}, so it is necessary to
estimate masses for the members of Upper Sco.

\begin{figure}
\epsscale{1.2}
\plotone{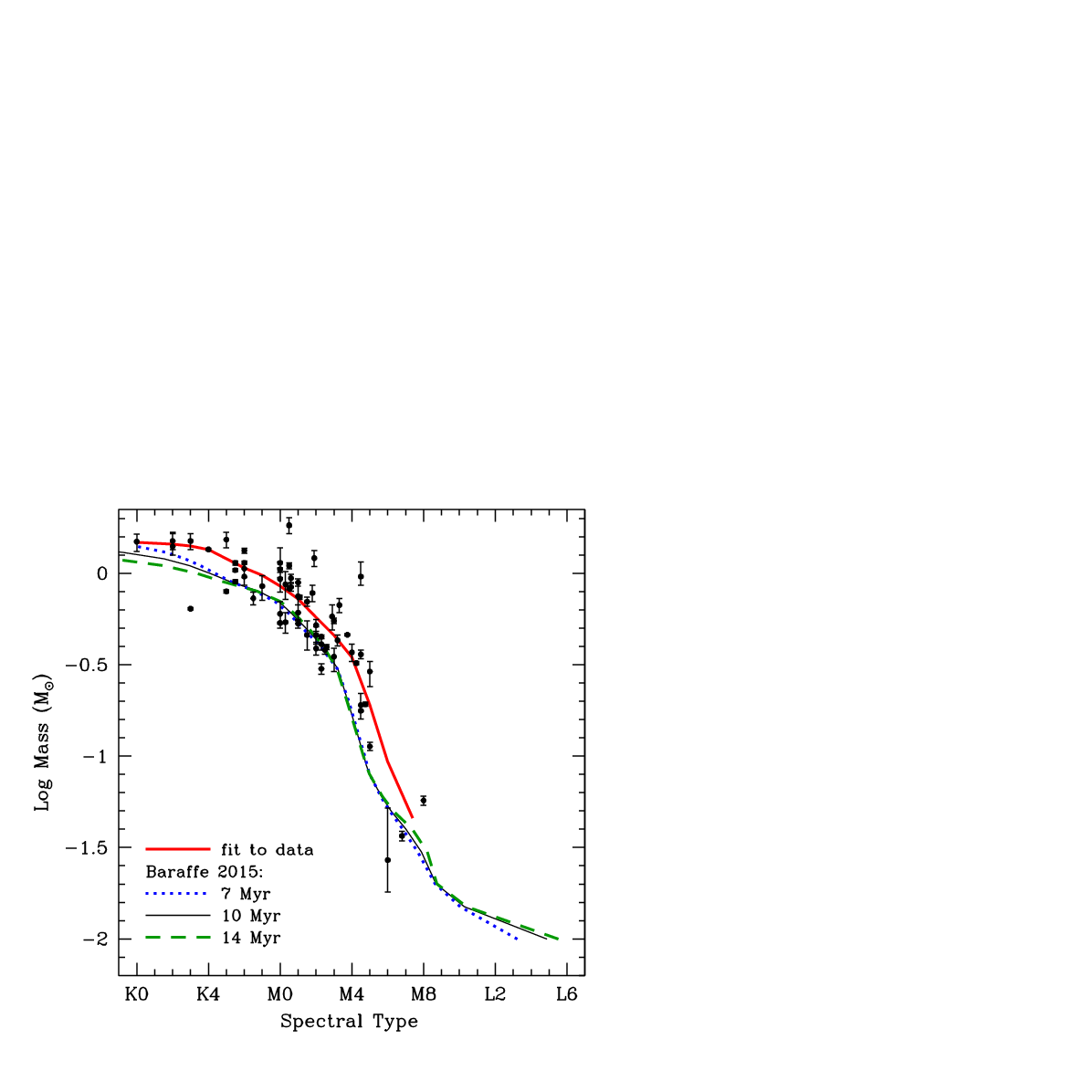}
\caption{Mass versus spectral type for young stars with
measurements of dynamical masses (Section~\ref{sec:mass}). 
I have included a fit to these data (Table~\ref{tab:fit}) and the relations 
predicted by evolutionary models for masses of 0.01--1.4~$M_\odot$ at ages 
of 7, 10, and 14 Myr \citep{bar15}.}
\label{fig:mass}
\end{figure}

Evolutionary models predict that young low-mass stars maintain roughly
constant effective temperatures for at least 10--20 Myr after 
the protostellar stage
\citep[e.g.,][]{bar15}, which is consistent with the fact that the peak of 
the IMF ($\sim$M5) occurs near the same spectral type for 
stellar populations across that
age range (e.g., L22). As a result, I have investigated the estimation of 
masses from spectral types in a way that is consistent with observational
constraints on masses of young stars. To do that, I have compiled
dynamical masses for young stars ($\lesssim30$~Myr) that are based on
binary orbits or rotation of circumstellar disks
\citep{gom09,sta14,cze15,dav16a,kra15,lod15,she19,sim19,bra21,peg21,sta22,fra23,tof23}.
In Figure~\ref{fig:mass}, I have plotted the dynamical masses 
as a function of spectral type for K and M stars. 
To focus on the more accurate measurements, those with relative errors
larger than 20\% are omitted.
Although its mass error exceeds the adopted threshold, I have included 
the companion PZ Tel B in Figure~\ref{fig:mass} because of its late spectral
type, where there are few measurements of dynamical masses.

Most of the spectral types in Figure~\ref{fig:mass} are those adopted by 
\citet{luh22sc} and \citet{luh23tau}, which apply to Sco-Cen and Taurus,
respectively. I have classified PZ Tel B as M6 through a comparison of the 
IR spectra from \citet{mai16} to young standards \citep{luh17}.
The components of the eclipsing binary system 2MASS J05352184$-$0546085 
\citep{sta06} lack resolved spectroscopy, so I have estimated their
spectral types by identifying the combination of standard spectra from
\citet{luh17} that best matches a SpeX spectrum of the system
(0.8--2.5~\micron, R=150) given the relative fluxes and temperatures of the
components \citep{gom09}. The resulting estimates are M8 and M6.8 for the
primary and secondary, respectively. In comparison, I find that the unresolved 
optical and IR spectra are best matched by M6.75 and M7, respectively.

The dynamical masses in Figure~\ref{fig:mass} exhibit a correlation with
spectral types, as expected. I have marked a fit to the median of that
sequence, which is defined in Table~\ref{tab:fit}. In this fit, the hydrogen 
burning mass limit \citep[0.075 $M_\odot$,][]{cha23} corresponds to a
spectral type of M6.4. To compare the correlation between dynamical masses and
spectral types to predictions of evolutionary models, I have converted
model temperatures to spectral types using the temperature scale from
\citet{her14} for types of $\leq$M8, adopting 2450 K for M9, and assuming
a decrement of 150 K for each subsequent subclass. The resulting temperature
scale is roughly consistent with temperatures estimated for late K and early M
stars in Taurus \citep{per24} and young objects at later types 
\citep{tre17,luh23}. The temperature scale for young L dwarfs is poorly
constrained, but the adopted scale is sufficient for showing the behavior
of the model isochrones in Figure~\ref{fig:mass}, and ultimately is not
used for my mass estimates in Upper Sco. In Figure~\ref{fig:mass},
I have plotted isochrones for masses of 0.01--1.4~$M_\odot$ at ages of 
7, 10, and 14 Myr \citep{bar15}, which span the ages of Upper Sco members 
(Section~\ref{sec:age}).
The isochrones are nearly identical for most of the masses in question,
which reflects vertical evolution in the H-R diagram. 
The isochrones are shifted relative to the data such that 
the models are predicting temperatures that are too high for a given
mass (assuming that the adopted temperature scale is accurate).

For objects in my IMF sample for Upper Sco that are between K0 and M7.4,
I have converted spectral types to masses using the fit to the sequence
of dynamical masses (Table~\ref{tab:fit}). Those masses agree on average
with the values derived from $M_K$ using the models of \citet{bar15}.
Therefore, for members later than M7.4, I have estimated masses from $M_K$ 
in the following manner.
For each of those late-type members, I have adopted the parallactic distance
from Gaia DR3 \citep{bai21} if $\sigma_{\pi}/\pi<0.1$. Otherwise, I have
assumed a distance of 145 pc, which is near the medial value for Upper Sco.
The $K_s$ photometry is corrected for extinction, which ranges from 
$A_K=0$--0.2 mag for most members \citep{luh20,luh22sc}. 
I have derived a fit to the median of the sequence
of Upper Sco members in $M_K$ versus spectral type. 
At late types, that fit should represent the median sequences at all ages
in Upper Sco since model isochrones across that age range
converge and overlap at spectral types later than M8.
At each spectral type, I have converted the value of $M_K$ on the fit 
to masses for ages of 7, 10, and 14 Myr using the models of \citet{bar15} 
and \citet{cha23} above and below 0.015 $M_\odot$, respectively. 
Those three mass estimates are assigned to all objects at the spectral 
type in question. Each object is added to the IMF with a weight of 1/3
at each of the three masses.

\subsection{Comparison of Upper Sco to Other Populations}

\begin{figure}
\epsscale{1.2}
\plotone{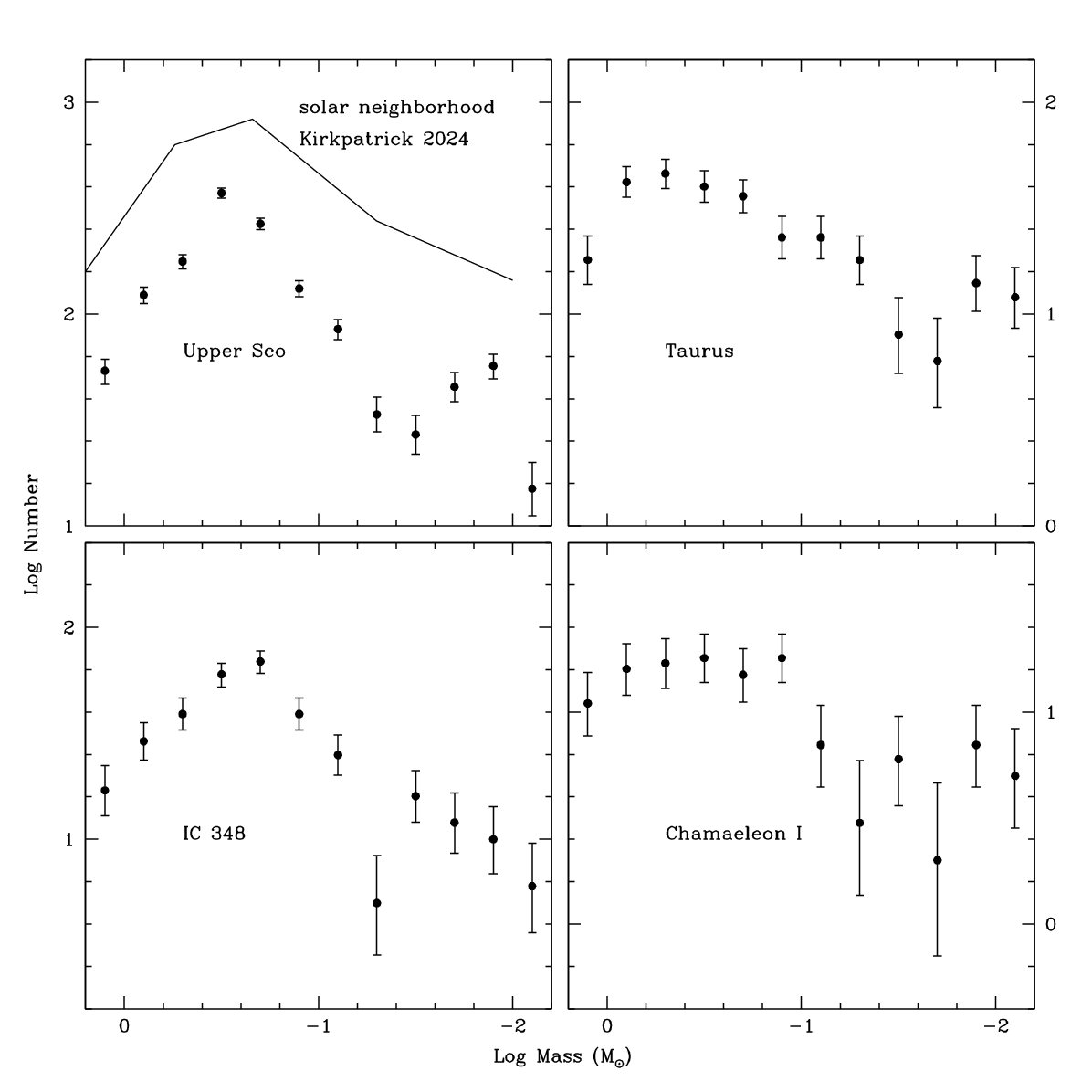}
\caption{IMFs for Upper Sco, Taurus, IC 348, Chamaeleon I, and the
solar neighborhood. The normalization for the latter is arbitrary.
The completeness limits for the first four IMFs are near 0.01~$M_\odot$.}
\label{fig:imf}
\end{figure}

My estimate of the IMF in Upper Sco is presented in Figure~\ref{fig:imf}.
The Salpeter slope is 1.35 \citep{sal55}
with the logarithmic mass bins that I have selected.
For comparison, I have included IMFs for three other star-forming
regions that have spectroscopic samples of members with well-defined
completeness limits that reach $\sim$0.01~$M_\odot$, consisting
of Taurus, IC 348, and Chameleon I \citep{esp17,esp19,luh16}.
I have derived the IMFs for those regions in the same manner as the
mass function for Upper Sco. The sample for IC 348 has been updated to 
include a few new members \citep{all20,luh20ic}. I have adopted 
ages of 3 Myr for Taurus and Chamaeleon I and 5 Myr for IC 348 
\citep{luh23tau,luh24}. In these IMFs, components of multiple systems
that have resolved spectroscopy are counted separately. Since only
very wide companions are resolved at the distances of these regions, 
the IMFs apply approximately to primary stars. The statistical errors in 
Figure~\ref{fig:imf} are based on the work by \citet{geh86}.

The IMF for Upper Sco exhibits a peak near 0.3~$M_\odot$, declines to lower
masses across the hydrogen burning limit, and rises somewhat below
0.03~$M_\odot$. Dips are also present in the substellar IMFs for
IC 348 and Taurus, although the one in Taurus occurs at lower masses. 
This structure in the substellar IMF may not be real, and instead may
reflect errors in the evolutionary models and the adopted relation 
between mass and spectral type across the hydrogen burning limit, where few 
dynamical masses are available.
I do not quote a value for the slope of the substellar IMF in Upper Sco
since it is sensitive to such errors as well as the selected mass range.
However, Figure~\ref{fig:imf} suggests that the substellar IMF is roughly
flat down to the 0.01~$M_\odot$ in Upper Sco and the other three regions. 

Spectroscopic samples of brown dwarfs reaching down to masses of 
$\sim0.01$~$M_\odot$ also have been obtained in the $\sigma$ Ori cluster 
\citep{pen12,zap17,dam23}. 
The resulting substellar IMFs appear be roughly consistent with the
mass functions in Figure~\ref{fig:imf}, although some of the objects in the 
$\sigma$ Ori samples lack spectroscopic confirmation of youth, and a 
close comparison of IMFs would require mass estimates for $\sigma$ Ori 
with the same methods applied in this study.

In Figure~\ref{fig:imf}, I have included a recent estimate for the IMF of
stars and brown dwarfs in the solar neighborhood \citep{kir24} (see also
\citet{bes24}), which has a completeness limit of 0.025~$M_\odot$ and extends 
to 0.01~$M_\odot$. Known components of multiple systems were counted 
separately in that IMF. Since multiplicity constraints are much better 
for nearby stars ($<$20 pc) than for members of star-forming regions
($>$140 pc), the IMF of individual objects from \citet{kir24} may not
be fully appropriate for a comparison to IMFs in ($\approx$primary)
star forming regions.
Nevertheless, the mass function in the solar neighborhood is broadly similar 
to the IMFs for the star-forming regions in Figure~\ref{fig:imf}.
The primary difference is the trough in the substellar IMF for Upper Sco,
and possibly the other young regions, which contributes to a difference
in the relative numbers of stars and brown dwarfs.
The number ratio of stars to brown dwarfs above 0.02~$M_\odot$ was 3.8
in the mass function from \citet{kir24} but is 1240/106 (12) 
and 287/33 (8.7) in the IMF samples for Upper Sco and IC 348, respectively.
Since the slope of the IMF is fairly steep across the hydrogen burning limit,
that ratio is sensitive to systematic errors in the mass estimates. 
For instance, the hydrogen burning limit corresponds to a spectral type 
of M6.4 in my fit to the dynamical masses in Figure~\ref{fig:mass}, 
but if it was earlier by one subclass, the number ratios of stars to brown 
dwarfs in Upper Sco and IC 348 would agree with that of the solar neighborhood.
The difference in the number ratios of stars to brown dwarfs
in this work and \citet{kir24} may also stem from a difference
between the mass functions of primary stars and individual objects.

\section{Circumstellar Disks}
\label{sec:disks}

Most of my adopted members of Upper Sco have been previously examined for
evidence of circumstellar disks in the form of IR excess emission 
\citep{esp18,luh20,luh22disks}. The exceptions consist of 57 of the newest
members \citep[][this work]{luh22sp,bou22}, nearly all of which are later 
than M6. Therefore, I present a disk analysis for the late-type members 
of Upper Sco.

Mid-IR images from WISE and the Spitzer Space Telescope \citep{wer04}
provide the best constraints on IR excess emission for members of Upper Sco. 
Spitzer offered greater sensitivity than WISE, but it observed only a small 
fraction of the association. Excess emission from a circumstellar disk is
larger at longer wavelengths, so data at longer wavelengths provide more
reliable detections of disks. For most of the new late-type members of 
Upper Sco, the W2 band of WISE/NEOWISE reaches the longest wavelengths among
the available data.
A few sources are also encompassed by Spitzer images longward of W2.
Three of the 57 new members that were absent from previous disk analysis lack
useful W2 photometry because of low S/N or blending with other stars.
For each $>$M6 member of Upper Sco, I have adopted W1 and W2 from either the
AllWISE Source Catalog or the CatWISE2020 Catalog, giving preference to the
measurements with smaller errors. These data are compiled in 
Table~\ref{tab:exc}.

\begin{figure}
\epsscale{1.2}
\plotone{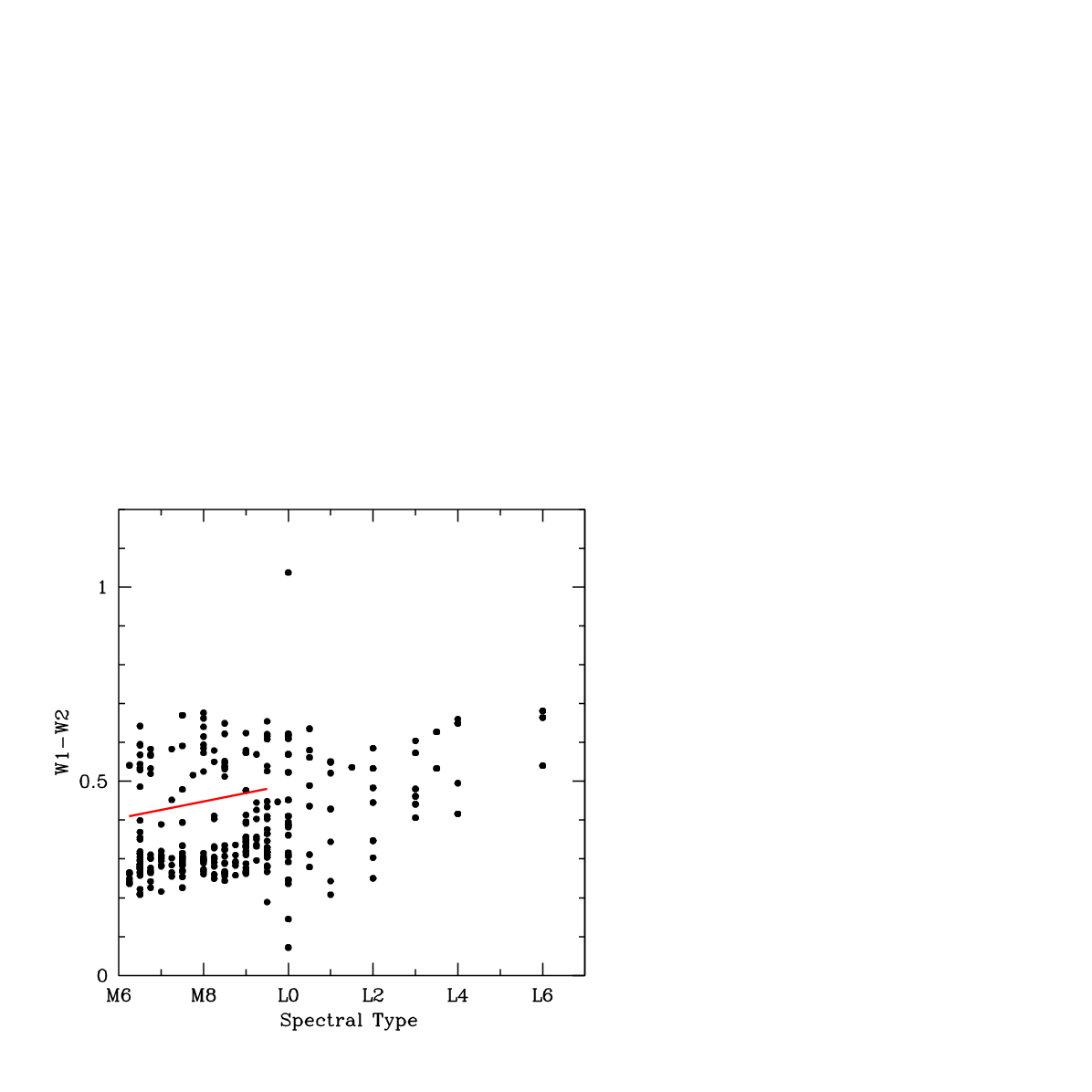}
\caption{W1$-$W2 versus spectral type for late-type members of Upper Sco.
At $<$L0, the sequence of photospheric colors is well-defined, and the
indicated threshold is used for identifying redder sources that have color
excesses (Table~\ref{tab:exc}). At later types, the photospheric sequence
is not well-defined, and only one member shows a significant excess.}
\label{fig:exc}
\end{figure}

Since the intrinsic photospheric colors of stars and brown dwarfs vary with 
spectral type, 
I have plotted W1$-$W2 versus spectral type in Figure~\ref{fig:exc}
for the late-type members of Upper Sco.  The data at M6.25--M9.5
exhibit a well-defined blue sequence that corresponds to stellar photospheres
and significantly redder sources that have W2 excesses.
I have marked a threshold in Figure~\ref{fig:exc} for identifying
M6.25--M9.5 sources that have W2 excesses. Table~\ref{tab:exc} includes a flag
for W2 excesses. The fraction of M6.25--M9.5 members with W2 excesses is 52/200,
which is similar to excess fractions from previous samples of Upper Sco 
members \citep{luh20}.

At types of $\geq$L0, the Upper Sco members form a single broad locus, 
which is likely a reflection of increasing W1$-$W2 with later L types
combined with larger uncertainties in the L spectral types compared to the
late M types (1--2 versus 0.5 subclass).
The spectral classifications are more uncertain for the L dwarfs because 
of a degeneracy between spectral type and reddening for low-resolution IR 
spectra \citep{luh17}.
As a result, a photospheric sequence is not clearly defined, and it is
difficult to reliably estimate photospheric color for any given object. 
Only one L-type object shows a noticeable color excess. 
Thus, W2 photometry provides poor constraints on the presence of disks for
the L-type members of Upper Sco. Data at longer wavelengths are needed to
detect disks among those objects.

\section{Conclusions}

I have sought to improve the completeness of the spectroscopic census
of brown dwarfs in the center of the Upper Sco association
and measure the substellar IMF down to masses of $\sim$0.01~$M_\odot$.
The results are summarized as follows:

\begin{enumerate}

\item
I have obtained IR spectra of 37 brown dwarf candidates in the center 
of Upper Sco. The spectra indicate that 35 of the candidates are young
and cool, making them likely members of the association.
This sample includes many of the faintest objects ($K=16$--17 mag) that 
have been confirmed as members through spectroscopy. 
Several of the targets have been observed spectroscopically in 
previous studies; the new spectra have higher S/Ns and broader wavelength
coverage, further refining the classifications.
In addition, I have analyzed new and archival spectra for 38 candidates
for low-mass stars in Upper Sco and other populations in Sco-Cen, most
of which exhibit evidence of youth in their spectra that is consistent
with membership.

\item
The populations within Upper Sco identified by \citet{mir22b}
span ages of 7--14 Myr based on a comparison to TWA and the adoption
of its expansion age of 10 Myr.

\item
I have compiled a catalog of 1753 sources that have spectral classifications,
evidence of youth, and astrometry that is consistent with membership in
Upper Sco.  Within the central field of Upper Sco defined
by \citet{luh20}, the new census of members has a high level of completeness 
down to $K=17$ mag, which corresponds to masses of $\sim0.007$--0.01~$M_\odot$
for the range of ages in Upper Sco according to evolutionary models.

\item
I have compiled measurements of dynamical masses of young stars
and brown dwarfs and used a fit to these data to estimate masses
for K0--M7 members of Upper Sco from their spectral types.
For cooler members at a given spectral type, I have estimated masses by
combining $M_K$ for the median of the Upper Sco sequence with the 
values predicted by evolutionary models for 7, 10, and 14 Myr.
These mass estimates have been used to construct an IMF for the central field
in Upper Sco. I have derived IMFs in the same manner for previous
membership samples in three other star-forming populations, consisting
of IC 348, Taurus, and Chameleon I.

\item
When using logarithmic mass bins (Salpeter slope = 1.35), the IMF for 
Upper Sco exhibits a peak near 0.3~$M_\odot$, declines to lower
masses across the hydrogen burning limit, and rises somewhat below
0.03~$M_\odot$. Similar troughs appear in some of the other young regions,
which may reflect errors in the evolutionary models and the adopted relation
between spectral type and mass.
Regardless of such errors, the data for Upper Sco and the other young
populations indicate that their substellar IMFs are roughly flat down to 
their completeness limits of $\sim$0.01~$M_\odot$.
These IMFs are broadly similar to mass functions recently measured
for the solar neighborhood \citep{bes24,kir24}, although the number ratio
of stars to brown dwarfs above 0.02~$M_\odot$ is higher
in the star-forming regions ($\sim10$ versus $\sim$4). That ratio in 
star-forming regions is sensitive to systematic errors in the mass estimates.
In addition, the mass functions in this work and \citet{kir24}
apply to primary stars and individual objects, respectively, which may
account for the different number ratios of stars to brown dwarfs.

\item
To search for evidence of disks among the new late-type members of 
Upper Sco found in recent studies, I have analyzed the W1$-$W2 colors for
all members later than M6. I find significant color excesses for
52 of the 200 members with spectral types of M6.25--M9.5, which is
similar to results from previous samples of members. 
For the L-type members, it is difficult to detect the small W2 excess emission
produced by typical disks around brown dwarfs because of the large 
uncertainties in spectral types, which preclude accurate estimates of the
photospheric colors. Reliable detections of disks among the
L-type members require data at longer wavelengths where excesses are larger.

\end{enumerate}

\begin{acknowledgments}

Gaia is a mission of the European Space Agency 
(\url{https://www.cosmos.esa.int/gaia}). Its data have been processed 
by the Gaia Data Processing and Analysis Consortium (DPAC, 
\url{https://www.cosmos.esa.int/web/gaia/dpac/consortium}). 
Funding for the DPAC has been provided by national institutions, in
particular the institutions participating in the Gaia Multilateral Agreement.
The IRTF is operated by the University of Hawaii under contract 80HQTR19D0030
with NASA. The COSMOS data were obtained through program 2022A-535205 at 
NOIRLab.  CTIO and NOIRLab are operated by the Association of Universities for
Research in Astronomy under a cooperative agreement with the NSF.
The Gemini data were obtained through programs 
GN-2022A-FT-109, GN-2022A-FT-204, GS-2022A-FT-206, and GN-2024A-Q-121.
Gemini Observatory is a program of NSF's NOIRLab, which is managed by the
Association of Universities for Research in Astronomy (AURA) under a
cooperative agreement with the National Science Foundation on behalf of the
Gemini Observatory partnership: the National Science Foundation (United States),
National Research Council (Canada), Agencia Nacional de Investigaci\'{o}n y
Desarrollo (Chile), Ministerio de Ciencia, Tecnolog\'{i}a e Innovaci\'{o}n
(Argentina), Minist\'{e}rio da Ci\^{e}ncia, Tecnologia, Inova\c{c}\~{o}es e
Comunica\c{c}\~{o}es (Brazil), and Korea Astronomy and Space Science Institute
(Republic of Korea). 2MASS is a joint project of the University of
Massachusetts and IPAC at Caltech, funded by NASA and the NSF.
WISE is a joint project of the University of California, Los Angeles,
and the JPL/Caltech, funded by NASA. 
The Center for Exoplanets and Habitable Worlds is supported by the
Pennsylvania State University, the Eberly College of Science, and the
Pennsylvania Space Grant Consortium.

\end{acknowledgments}

\clearpage

\begin{deluxetable}{llll}
\tabletypesize{\scriptsize}
\tablewidth{0pt}
\tablecaption{Summary of Spectroscopic Observations\label{tab:log}}
\tablehead{
\colhead{Telescope/Instrument} &
\colhead{Mode/Aperture} &
\colhead{Wavelengths/Resolution} & 
\colhead{Targets}}
\startdata
IRTF/SpeX & prism/$0\farcs8$ slit & 0.8--2.5~\micron/R=150 & 21 \\
CTIO 4~m/COSMOS & red VPH/$1\farcs2$ slit & 0.55--0.95~\micron/4 \AA & 13 \\
Gemini North/GNIRS & 31.7~l~mm$^{-1}$/$1\arcsec$ slit & 0.9--2.5~\micron/R=600 & 21 \\
Gemini South/FLAMINGOS-2 & JH/$1\farcs08$ slit & 0.9--1.8 \micron/R=400 & 5 \\
NTT/EFOSC & 300~l~mm$^{-1}$/$1\arcsec$ slit & 0.6--1 \micron/13 \AA & 15 
\enddata
\end{deluxetable}

\begin{deluxetable}{ll}
\tabletypesize{\scriptsize}
\tablewidth{0pt}
\tablecaption{Spectroscopic Data for Candidate Members of Sco-Cen\label{tab:spec}}
\tablehead{
\colhead{Column Label} &
\colhead{Description}}
\startdata
Gaia & Gaia DR3 source name \\
irname & Source name from UKIDSS or VISTA VHS\\
RAgdeg & Gaia DR3 right ascension (ICRS at Epoch 2016.0)\\
DEgdeg & Gaia DR3 declination (ICRS at Epoch 2016.0)\\
RAudeg & Right ascension from UKIDSS or VISTA VHS (J2000) \\
DEudeg & Declination from UKIDSS or VISTA VHS (J2000) \\
SpType & Spectral type\tablenotemark{a}\\
young & Young?\\
Instrument & Instrument for spectroscopy\\
Date & Date of spectroscopy
\enddata
\tablenotetext{1}{Uncertainties are 0.25 and 0.5~subclass for optical and
IR spectral types, respectively, unless indicated otherwise.}
\tablecomments{
The table is available in its entirety in machine-readable form.}
\end{deluxetable}

\begin{deluxetable}{ll}
\tabletypesize{\scriptsize}
\tablewidth{0pt}
\tablecaption{Adopted Members of Upper Sco with Spectral Classifications at $l=343$--$360\arcdeg$ and $b=10$--$30\arcdeg$\label{tab:usco}}
\tablehead{
\colhead{Column Label} &
\colhead{Description}}
\startdata
Gaia & Gaia EDR3 source name \\
UGCS & UKIDSS Galactic Clusters Survey source name \\
2MASS & 2MASS Point Source Catalog source name \\
Name & Other source name \\
RAdeg & Right ascension (ICRS)\\
DEdeg & Declination (ICRS)\\
Ref-Pos & Reference for right ascension and declination\tablenotemark{a} \\
SpType & Spectral type \\
r\_SpType & Spectral type reference\tablenotemark{b} \\
Adopt & Adopted spectral type 
\enddata
\tablenotetext{a}{Sources of the right ascension and declination are Gaia EDR3,
DR6 of VISTA VHS, DR10 of the UKIDSS Galactic Clusters Survey,
the 2MASS Point Source Catalog, and high-resolution
imaging \citep{ire11,kra14,lac15,bry16}.}
\tablenotetext{b}{
(1) \citet{hou88};
(2) \citet{can93};
(3) \citet{luh18};
(4) \citet{daw14};
(5) \citet{esp18};
(6) \citet{luh20};
(7) \citet{riz15};
(8) \citet{pre98};
(9) \citet{bou22};
(10) measured in this work with the spectrum from \cite{bou22};
(11) \citet{lod08};
(12) \citet{bon14};
(13) \citet{luh22sp};
(14) \citet{lod06};
(15) \citet{kun99};
(16) \citet{hil69};
(17) \citet{man20};
(18) \citet{pec16};
(19) \citet{pen16};
(20) this work;
(21) \citet{hou82};
(22) \citet{cor84};
(23) \citet{all13b};
(24) \citet{mar04};
(25) \citet{mar10};
(26) \citet{sle08};
(27) \citet{chi20};
(28) \citet{pre02};
(29) \citet{mor01};
(30) \citet{vie03};
(31) \citet{rei08};
(32) \citet{ard00};
(33) \citet{wal94};
(34) \citet{kir10};
(35) \citet{all13a};
(36) \citet{fah16};
(37) \citet{tor06};
(38) \citet{giz02};
(39) \citet{her14};
(40) \citet{sle06};
(41) \citet{ria06};
(42) \citet{bes17};
(43) \citet{lod18};
(44) \citet{kra15};
(45) \citet{cod17};
(46) \citet{laf11};
(47) \citet{lac15};
(48) \citet{dav19b};
(49) \citet{ans16};
(50) \citet{pra02};
(51) \citet{pre01};
(52) \citet{alm24};
(53) \citet{cru03};
(54) \citet{bej08};
(55) \citet{her09};
(56) \citet{lod21};
(57) \citet{kra07};
(58) \citet{lod11a};
(59) \citet{mue11};
(60) \citet{kra09};
(61) \citet{pec12};
(62) \citet{bil11};
(63) \citet{laf08};
(64) \citet{luh17};
(65) \citet{man16};
(66) \citet{dav16b};
(67) \citet{sta17};
(68) \citet{sta18};
(69) \citet{coh79};
(70) \citet{pra03};
(71) \citet{eis05};
(72) \citet{mur69};
(73) \citet{mar98a};
(74) \citet{pra07};
(75) \citet{mcc10};
(76) \citet{lod15};
(77) \citet{dav16a};
(78) \citet{car06};
(79) \citet{luh05usco};
(80) \citet{mar98b};
(81) \citet{esp20};
(82) \citet{all20};
(83) \citet{cie10};
(84) \citet{bow11};
(85) \citet{bow14};
(86) \citet{bra97};
(87) \citet{bou92};
(88) \citet{bow17}.}
\tablecomments{
The table is available in its entirety in machine-readable form.}
\end{deluxetable}

\begin{deluxetable}{ll}
\tabletypesize{\scriptsize}
\tablecaption{Fit to Median of Dynamical Mass versus Spectral type\label{tab:fit}}
\tablehead{
\colhead{Spectral Type} &
\colhead{log M}\\
\colhead{} &
\colhead{($M_\odot$)}}
\startdata
K0 &   0.17 \\
K2 &   0.16 \\
K3 &   0.15 \\
K4 &   0.13 \\
K5 &   0.08 \\
K6 &   0.03 \\
K7 &  -0.01 \\
M0 &  -0.07 \\
M1 &  -0.14 \\
M2 &  -0.24 \\
M3 &  -0.34 \\
M4 &  -0.46 \\
M5 &  -0.72 \\
M6 &  -1.03 \\
M7.4 & -1.34
\enddata
\end{deluxetable}

\begin{deluxetable}{ll}
\tabletypesize{\scriptsize}
\tablewidth{0pt}
\tablecaption{W2 Excess Flags for $>$M6 Members of Upper Sco\label{tab:exc}}
\tablehead{
\colhead{Column Label} &
\colhead{Description}}
\startdata
Gaia & Gaia EDR3 source name \\
UGCS & UKIDSS Galactic Clusters Survey source name \\
2MASS & 2MASS Point Source Catalog source name \\
Name & Other source name \\
Ref-Pos & Reference for right ascension and declination\tablenotemark{a} \\
W1mag & WISE W1 magnitude \\
e\_W1mag & Error in W1mag \\
W2mag & WISE W2 magnitude \\
e\_W2mag & Error in W2mag \\
exc & Excess present in W2?
\enddata
\tablenotetext{a}{Sources of the right ascension and declination are Gaia EDR3,
DR6 of VISTA VHS, DR10 of the UKIDSS Galactic Clusters Survey, and
the 2MASS Point Source Catalog.}
\tablecomments{
The table is available in its entirety in machine-readable form.}
\end{deluxetable}

\clearpage

\end{document}